# IMPACT OF CULTURE ON THE ADOPTION OF DIABETES SELF-MANAGEMENT APPLICATIONS: CAPE FLATS, SOUTH AFRICA

Fazlyn Petersen, University of the Western Cape, Fapetersen@uwc.ac.za

**Abstract:** Diabetes is a global health problem with a high mortality rate. The research indicates low levels of technology use amongst diabetic patients in low socioeconomic environments and minority groups. We posit that the culture of patients is a potential reason for the low adoption and use of technology. However, research on the proliferation of culture at an individual level is limited. Therefore, this paper assessed the influence of culture on mobile application adoption and use amongst diabetic patients in the Cape Flats, South Africa. This study used key constructs from the Theory of Planned Behaviour (TPB) and Hofstede's cultural dimensions. It was analysed using survey data from 439 respondents using purposive sampling. It was found that the dimensions of Hofstede and the Theory of Planned Behaviour can identify how culture influences mobile application adoption of diabetic patients in the geographical Cape Flats area. However, this research indicates a stronger relationship between culture and diabetes self-management activities than culture and the adoption of mobile applications.

**Keywords:** Mobile applications, Theory of Planned Behaviour (TPB), Hofstede's cultural dimensions, Western Cape, diabetes self-management

## 1. INTRODUCTION

Non-communicable disease (NCD) is the leading cause of death globally (Mutyambizi, Booysen, Stokes, Pavlova, & Groot, 2019) and it is expected to surpass other causes of mortality in Africa by 2030 (World Health Organization, 2014). Diabetes mellitus, a worldwide pandemic presents one of the major non-communicable diseases (NCDs) in South Africa. Diabetes is a chronic disease with a high level of mortality rate and has become increasingly prevalent in low-middle income countries (LMICs) (World Health Organization, 2016). In Africa, it is estimated that 16 million people are living with diabetes (International Diabetes Federation, 2017) and it is expected to surpass other causes of mortality in Africa by the year 2030 (World Health Organization, 2014). Statistics show that in 2017, 10.4% of the Western Cape population has succumbed to diabetes (Statistics South Africa, 2016).

Self-management is an essential part of diabetes management. diabetic patients must follow the self-management activities consist of healthy eating, being active, monitoring, taking prescribed medication, problem-solving, healthy coping and reducing risks (American Association of Diabetes Educators, 1997). Poor self-management can lead to significant mortality and a poor quality of life (Reyes, Tripp-Reimer, Parker, Muller, & Laroche, 2017). The risk of having diabetes is determined by factors such as age, ethnicity, socioeconomic status and lifestyle factors (World Health Organization, 2016). Although the pervasiveness of diabetes varies with socioeconomic status, the disparities can be worsened by the unhealthy lifestyles adopted by individuals (Mukong, Van Walbeek, & Ross, 2017).

In the current era, there is significant potential for modern technology, such as mobile technologies to be used to address disease management. Mobile health (m-health) interventions seem to be developed and implemented in a sociocultural vacuum – *"the template for many m-health interventions are mainly interventions from developed countries"* (Müller, 2016 p.295). Therefore, understanding patients self-management and cultural value systems is an important factor for designing effective self-management interventions that could ultimately influence self-management behaviours (Ayele, Tesfa, Abebe, Tilahun, & Girma, 2012). Designing ICT interventions must consider a *"patient-centred communication style that incorporates patient preferences; assesses literacy and numeracy, and addresses cultural barriers to care"* (American Diabetes Association, 2015 p.S5) to improve health outcomes.

Culture is defined as *"the collective programming of the mind which distinguishes the members of one human group from another"* (Hofstede, Hofstede & Minkov, 2010 p.5). Culture is context-specific. Research indicated that culture is unique to geography and may not be seen in isolation (Dwivedi, Shareef, Simintiras, Lal, & Weerakkody, 2016). The culture within the South African context is complex due to apartheid and the division of the population based on race and gender (Bekker, Leildé, Cornelissen, & Horstmeier, 2000; Shefer et al., 2008). The Western Cape is heterogeneous in terms of culture, linguistics and religion (Bekker et al., 2000). The Cape Flats, an area in the Western Cape, was created during Apartheid to move non-white residents out of the City Centre (Farrar, Falake, Mebaley, Moya, & Rudolph, 2019). The Cape Flats area consisted of low-cost public housing known as townships and informal settlements. (Farrar et al., 2019). The Cape Flats is home to "*predominantly isiXhosa-speaking 'Black Africans' and people belonging to an ethnically heterogeneous group of brown people known colloquially as 'Coloureds'*" (Farrar et al., 2019 p.3). The research indicates that in a 'black' and 'coloured' community in the Western Cape there are traditional roles of male dominance and female subservience however this is changing over time (Bekker et al., 2000). There is also a growing diabetes population in Cape Town (Kengne & Sayed, 2017). Diabetes control is lower in patients with low socioeconomic conditions, residing in areas such as Bishop Lavis (Booysen & Schlemmer, 2015) and Khayelitsha (Guwatudde et al., 2018) in the Cape Flats.

There has been extensive research about the link between cultural factors and the acceptance of technology (Al-jumeily & Hussain, 2014; Barton, 2010; Kovačić, 2005; Srite & Karahanna, 2006; Tarhini, Hone, Liu, & Tarhini, 2017). These studies illustrate that cultural backgrounds play an important role in affecting the acceptance and use of technology. Researchers have applied Hofstede's cultural dimensions in technology acceptance and adoption models (Caporarello, Magni, & Pennarola, 2014; Hoque & Bao, 2015). However, there is a low level of technology acceptance and use for diabetes self-management in geographical areas, such as the Cape Flats (Petersen, Pather, & Tucker, 2018). Additionally, the effect of culture on technology adoption has not yet been studied in this context. Therefore, this research objective determined how culture influence m-health adoption for self-management amongst diabetic patients in this context.

## 2. CULTURE AND ADOPTION

Hofstede's cultural dimensions are defined and applied to South Africa in Table 1 below:

**Table 1 South African culture based on Hofstede's cultural dimensions**

| Construct | Definition based on Hofstede et al. (2010) | South African cultural dimensions |
|---|---|---|
| **Power distance (PD)** | The *"extent to which the less powerful members of institutions and organizations within a country expect and accept that power is distributed unequally"* (p.61). | While Hofstede's cultural dimensions are indicated for South Africans (Cronjé, 2006), in some research, the following reference (Hofstede, 2019) indicates that it is for White South Africans. It also cautions that these values may not apply to the Black or Coloured South African population. The PD value for White South Africans indicates that they accept a hierarchical order where power is not equally distributed (Hofstede, 2019). |
| **Individualism-collectivism (IC)** | Individualism refers to *"societies in which everyone is expected to look after him- or herself and immediate family.* Collectivism refers to *"societies in which people from birth onward are integrated into strong, cohesive in-groups"* (p.92). | The IC value for White South Africans indicates that they prefer taking care of themselves and their immediate families only (Hofstede, 2017). This value may be different for closely-knit Black or Coloured families and communities. |
| **Masculinity-femininity** – | Masculinity stands for a *"society in which emotional gender roles are distinct"* (p.519). Femininity is seen as a *"society in which emotional gender roles overlap: both men and* | White South Africans view South Africa as a masculine society where society is driven by achievement and success (Hofstede, 2019). |
| | *women are supposed to be modest, tender, and concerned with the quality of life"* (p.517). | |
| **Uncertainty avoidance (UA) –** | *"The extent to which the members of a culture feel threatened by ambiguous or unknown situations"* (p.191). | There is a low UA value for White South Africans. The value is indicative of a society that is more relaxed and does not fear uncertainty (Hofstede, 2019). |
| **Long-term orientation (LTO)** | *"The fostering of virtues oriented toward future rewards—in particular, perseverance and thrift"* (p.239). | White South Africans indicate a low score. Therefore, they value traditions and may view change negatively (Hofstede, 2019). |
| **Indulgence-Restraint (IR)** | Indulgence refers to *"a society that allows relatively free gratification of basic and natural human desires related to enjoying life and having fun"* (p.519). Restraint refers to a *"society that suppresses gratification of needs and regulates it using strict social norms"* (p.521). | The high IR value for White South Africans indicates that they prefer acting on their impulses, enjoy having fun and spending as they wish (Hofstede, 2017). |

Research also indicates several potential culture-related influences on technology adoption, which warrants further investigation (Petersen, Brown, Pather, & Tucker, 2019). Research shows that differences between social structures, the standard of living and religion may influence individuals obligations to behave healthily (Hjelm, Bard, Nyberg, & Apelqvist, 2003). The research indicates that culture may impact diabetes self-management. For example, individuals from different societies practice diverse religious obligations, eating habits social customs and this may influence their beliefs, behaviour, perception and attitudes towards health (Swierad, Vartanian, & King, 2017). Research shows that culture and socioeconomic status shaped diabetic patient eating patterns (Matima, Murphy, Levitt, BeLue, & Oni, 2018).

Cultural values influencing technology adoption have been widely studied from a national perspective, the studies on an individual level are limited (Sunny, Patrick, & Rob, 2019). Research into the role of culture on diabetic patients' adoption of m-health is warranted given the previous studies have shown that behavioural intention on its own is not a predictor of use (Petersen et al., 2018). Additionally, other factors that influence m-health adoption and use such as culture and research should be centred on how people use technology (Müller, 2016).

The four prominent models for adoption used in mobile health services are the Theory of Reasoned Action (TRA) (M. . Fishbein & Ajzen, 1975), Technology Acceptance Model (TAM) (Davis, 1989), the Theory of Planned Behavior (TPB) (M. . Fishbein & Ajzen, 1975) and the Unified Theory of Acceptance and Use of Technology (UTAUT) (Venkatesh, Morris, Davis, & Davis, 2003) are summarised in Figure 1.

TRA includes the key beliefs formulating attitude in that the attitude toward the behaviour and normative beliefs are important in determining whether individuals will perform the desired behaviour (Fishbein & Ajzen, 1975). TRA includes the key beliefs formulating attitude in that the attitude toward the behaviour and normative beliefs are important in determining whether individuals will perform the desired behaviour (Fishbein & Ajzen, 1975). The desired behaviour in this study refers to the performance of self-management activities.

The TPB model includes expands TRA by adding perceived behavioural control (PBC). PBC is the "perceived ease or difficulty of performing the behaviour" (Ajzen, 1991 p.188). In a comparison between TRA and TPB, the inclusion of PBC explained more of the variance in behaviour than TRA (Madden, Ellen & Ajzen, 1992). TPB is used more extensively in predicting health behaviours (LaMorte, 2018).

TAM compares favourably with alternative models such as TRA and TPB (Venkatesh & Davis, 2000). Based on Davis (1989), TAM includes perceived usefulness (whether patients believe that using a mobile application would improve their health) and perceived ease of use (whether patients believe that using a mobile application would be effortless).

The UTAUT model expands TAM with the inclusion of Social Influence (SI): "the degree to which an individual perceives that important others believe he or she should use the new system" (Venkatesh *et al.,* 2003:451) and Facilitating Conditions (FC): "the degree to which an individual believes that an organisational and technical infrastructure exists to support the use of the system" (Venkatesh *et al.,* 2003:453).

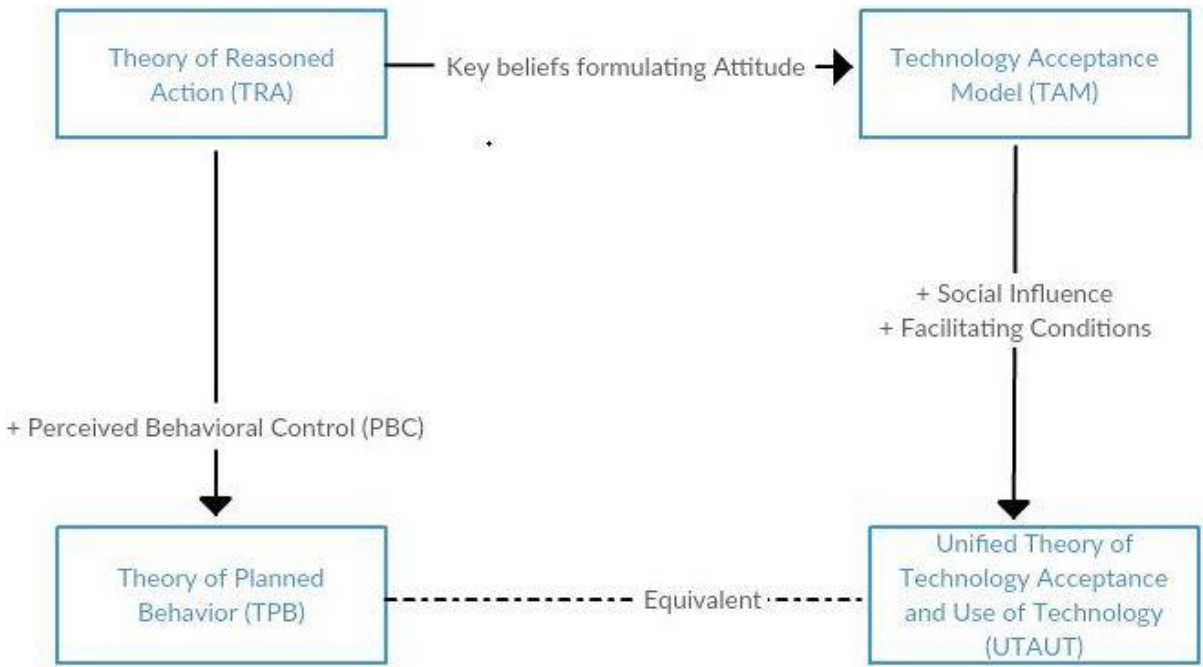

**Figure 1 Summary of technology adoption models (based on** (Sun, Wang, Guo, & Peng, 2013)**).**

UTAUT and TPB constructs cover similar constructs in terms of mobile health services (Sun et al., 2013). However, TPB has been used in studies to predict self-management behaviours in people who are at risk of diabetes and diagnosed with type 2 diabetes (Boudreau & Godin, 2014; Traina, Mathias, Colwell, Crosby, & Abraham, 2016). Additionally, it has been found that the TPB constructs are significant predictors to carry out self-management behaviours of diabetic patients (Gatt & Sammut, 2008). Research shows that the TPB has been used to attain an improved knowledge of the perceptions underlying the adoption of health-related behaviours (Akbar, Anderson, & Gallegos, 2015). UTAUT was not effective in predicting the acceptance and use of ICT for diabetic patients in the Western Cape (Petersen et al., 2018) therefore, TPB was selected as the basis of the research model.

## 3. RESEARCH MODEL

In this study context, culture is viewed based on diabetic patients' perceptions of health and adoption of mobile applications. To address the research objective, the research model was developed consist of two theoretical frameworks. Hofstede cultural dimension and TPB was used (Figure 2).

The TPB constructs are imperative predictors of *"intent to carry out self-management behaviour in persons with type 2 diabetes"* (Gatt & Sammut, 2008 p.1525). Moreover, all of the Hofstede constructs can influence technology acceptance and use. The TPB model can predict diabetes patients' self-management behaviours and thus can affect the acceptance and use of m-health applications. The TPB model can be used as predictors of the Hofstede cultural model as diabetes patients' attitude and values can be influenced by their beliefs regarding whether to adopt m-health applications.

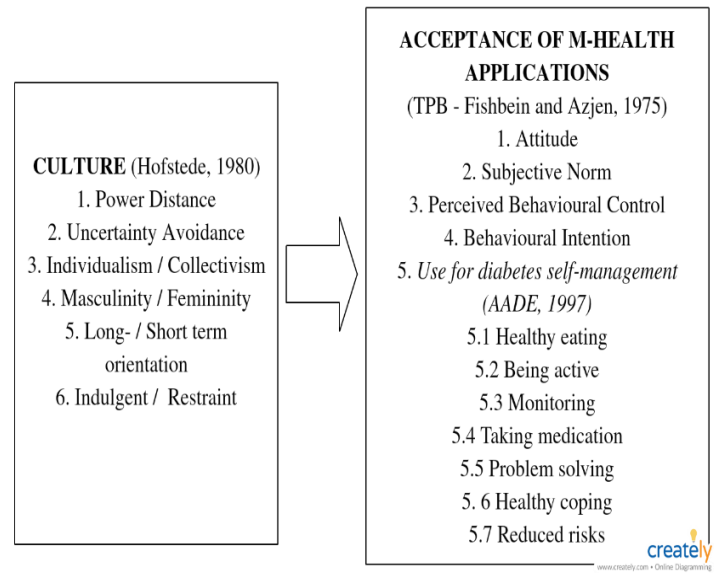

**Figure 2 Research model**

In summary, the research model (Figure 2) used two main elements used to gain insights into diabetes patients' adoption of m-health applications for self-management activities:
1. Hofstede's cultural dimensions model was used to understand whether diabetes patients' culture, influences their self-management behaviours and whether their culture influences their mobile application adoption for their self-management.
2. Theory of Planned Behaviour (TPB) was used to provide insight into respondents' adoption of m-health applications to manage their diabetes condition.

The inclusion of the AADE 7 self-management behaviour activities model was used to understand diabetic patients' self-management behaviours.

## 4. RESEARCH DESIGN AND METHODOLOGY

This research paradigm used was interpretivism. Interpretivism is associated with an inductive approach. An inductive approach was used for understanding the impact of culture on patients' adoption of diabetes self-management applications (Saunders, Lewis, & Thornhill, 2009).

### 4.1. Data collection

Qualitative data was collected from 438 diabetic patients via a survey. Purposive sampling was used and respondents eligible for inclusion were diabetic patients residing in the Cape Flats. The areas that are predominant in this study are from Mitchell's Plain and Khayelitsha as these areas showed a low level of diabetes self-management was practised (Daivadanam et al., 2017; Khalied, 2017).

### 4.2. Data Analysis

The study used thematic content analysis to identify how culture influence mobile application adoption and use for diabetes self-management. The data analysis was based on the key thematic areas identified in the research model (Figure 2). The qualitative evidence was analysed with Atlas.ti software data to identify how culture influences mobile health adoption. The themes were derived from the core constructs of the TPB (Davis, 1989), Hofstede cultural dimensions and AADE 7 self-management behaviour activities. Respondents' responses were coded and grouped according to the constructs and thereafter patterns and relationships have been identified based on existing research.

### 4.3. Ethical considerations

The research was conducted according to the ethical standards specified by the Higher Committees at the University of the Western Cape. Respondents received an information sheet and consent form before any data being collected. The respondents' personal information was protected as no clinical data or unique identifiers such as respondents' ID numbers were collected. Respondents' were informed that this was a voluntary process. Respondents reserved the right to refuse to answer any question with which they felt uncomfortable.

## 5. FINDINGS

The themes identified in this study aimed to identify how culture influences mobile application adoption of diabetes self-management. The section commences with the respondents' demographics and concludes with thematic content analysis using the research model in Figure 2.

### 5.1. Respondent demographics

The analysis of demographic data indicated that the majority of respondents were female (58.6%), with grade 12 as their highest level of education (36.2%). Respondents were older than 50 (39.9%), with type 2 insulin-resistant diabetes, using oral diabetes medication such as Metformin or Glucophage. Respondents resided in the Cape Flats in the City of Cape Town. The top five areas were Belhar (11.9%), Athlone (9.7%), Mitchell's Plain (8.6%), Khayelitsha (4.7%) and Gugulethu (3.5%). The most important observation is that the majority (64.7%) of respondents strongly disagreed that culture affected their ability to adopt diabetes mobile applications.

### 5.2. Thematic Content Analysis

When a diabetic patient trusts their doctors or prefers visiting a doctor as opposed to using an m-health application, they form part of a power distance society. This was evidenced by the following quote, *"I don't use mobile applications at all for my diabetes as I'm high risk and prefer professional advice and assistance"* (Coloured, male, older than 50 years). The quotation also shows the relationship between power distance, reducing diabetes risks self-management activity and the use of mobile applications.

When a diabetic patient subscribes to a high uncertainty avoidance society, s/he may find it difficult to adjust to technological changes. Uncertainty avoidance was evidenced by the following quote, *"Nobody I know uses it. Most of my family has and they never used it. We just used to the old way of going to the doctor"* (Coloured, male, 18 - 24 years). The comment also highlights that uncertainty avoidance may not be influenced by age as this respondent was young. The quote also links to a low long-term orientation.

In terms of individualism, if a patient forms part of an individualistic society, s/he will make their own informed decision as to how to manage their condition. Individualism was evidenced as respondents indicated that others do not influence their decision to use m-health applications. *"My culture does not affect my ability to use mobile health application to manage my diabetes. This is mainly because I do not centre my life based on what other people may say or think because my health is important to me"* (Black, female, 25 – 34 years).

A link between individualism and perceived behavioural control is evidenced by the following quote, *"I am allowed to use any smart devices, and the problems is that I find technology difficult for me"* (Black, female, 35 - 49 years).

There was also a link between individualism and the problem-solving diabetes self-management activity, *"As time has passed, our culture (isiXhosa) has become more flexible to the fact that we use technology for innovation and to solve problems so my culture does not restrict me to gain access to proper technological solutions for illness or treatment"* (Black, male, 25 – 34 years).

However, a collectivist cultural construct relates to the subject norm theme in TBP as family and important others influences' the decision to adopt mobile applications. Patients who form part of collectivistic cultures will make health-related decisions based on the values and beliefs of their societies. A respondent highlighted this point in the statement, *"It does not really affect my culture as most of the people that I am surrounded by has mobile applications and are technologically advanced. Therefore they largely influence my ability into using mobile applications"* (Coloured, male, 25 – 34 years).

Respondents indicated a relationship between collectivism, social norms and eating healthy. A negative relationship is evidenced by the following quotes, *"In our Black culture we have gotten used to the whole notion of eating full meal plates every day and that does affect my health"* (Black, female, 35 - 49 years).

However, a positive relationship was also indicated, *"They know my diet, so they always prepare the right food for my situation like fruit and vegetables* (Black, male, 25 – 34 years).

When a patient forms part of the feminine society, s/he will care for others and make informed health decisions to assist themselves and others in leading healthier lifestyles. The femininity cultural construct was evidenced by the following quote, *"As a single parent to three daughters' majority of my day consists of praying, cooking, cleaning and family time. I, therefore, don't have much time to figure out how or remember to use mobile applications"* (Coloured, female, older than 50 years). Respondents indicated that they do not have time to manage their condition or use mobile applications due to their family responsibilities. Hence, there is also a link to social norms.

A link between femininity and a negative attitude to using m-health applications was also highlighted by the following statement, "*I live in a township and mobile applications aren't really something that people do. I am also a 56 year old woman and I don't really have the energy to try and learn how these fancy are used"* (Black, female, older than 50 years).

Respondents indicated that there was a low long-term orientation. They are more traditional, they prefer going to the doctor than using an application. Evidence from respondents indicated that *"My culture does not affect my ability to use mobile applications. I just feel it's better to do things the old-fashioned way"* (Black, male, 25 – 34 years).

Adhering to religious practices was also indicated and linked to the self-management activity of eating healthy, *"During the month of Ramadaan, we are not to eat anything during the day. This obviously means I cannot have consistent meals through the day as we cannot be fasting and eating etc. This does affect my energy levels and allow for room of fatigue to come about"* (Coloured, female, older than 50 years).

The following quotation indicated a link between the low long-term orientation and perceived behavioural control*, "I am not exposed to technology so I can't use some technological tools that help me manage my illness"* (Black, female, 25 – 34 years).

Long-term orientation and adhering to religious practices were also linked to taking prescribed medication. This is evidenced by the following quotations, *"I go to a traditional healer twice a month to for herbs to keep my body clean and strong"* (Black, female, 25 - 34 years) and *"It teaches discipline and that has helped in terms of me taking my medication every time, and not eating what I am not supposed to eat"* (Coloured, female, 35 - 49 years).

However, a short-term orientation was evidenced by the following quote, *"My culture encourage change and adapting to change. So it encourages technology"* (Coloured, male, older than 50 years).
When a diabetic patient subscribes to an indulgence society, s/he will make health-related decisions that are satisfactory to them to ensure that they are happy. Respondents indicated that they prefer spending money on things that are fulfilling to them. The view of indulgence is supported by the following quote, *"I can do what I what as long it helps me to have a long and prosperous life"* (Coloured, male, 35 - 49 years).

The link between an indulgence society and healthy coping was supported by the following quote, *"My culture supports me in adopting habits which will allow me to live a full and healthy life"* (Coloured, female, 35 - 49 years).

As indicated in the demographic analysis, the majority of respondents (64.7%) indicated that culture did not impact their adoption of diabetes self-management applications. Socioeconomic status was emphasised by the following quote, *"I would not exactly say my culture affects it rather my economic status as a domestic worker. I am a single mother of 4 and grandmother of 2 all depending on me. I cannot afford a smartphone and therefore not so familiarised with the internet and mobile applications"* (Black, female, 35 - 49 years). However, this quote also indicates social norms and being part of a feminine society.

Respondents indicated a lack of knowledge affects the use of diabetes self-management applications, "*Culture does not forbid it but there is a lack of knowledge* (Coloured, female, older than 50 years). Age and affordability were identified as important factors that negatively affect the use of applications, *"Not interested* *in using a smart phone as my age doesn't allow for it. [I] don't have enough funding for data or the phone"* (Coloured, female, older than 50 years). The crime was also highlighted *"The area that I live in (Bridgetown) is not safe and I have a fear of being robbed if I am using my smartphone in public"* (Coloured, male, 18 – 24 years).

## 6. DISCUSSION

The evidence did not indicate that culture influences the adoption of diabetes self-management applications in this context, unlike the existing research. However, this research indicated a stronger relationship between culture and diabetes self-management activities than culture and the adoption of mobile applications. The finding that culture influences healthy eating especially for Coloured and Black respondents is supported by research that indicates culture affects diabetic patient eating patterns (Matima et al., 2018). The link between culture and exercise and the use of diabetes mobile applications for exercise was not explicitly found in this study.

Cultural, social and family influences can shape people's beliefs and attitudes. Other factors such as age, affordability and a lack of knowledge also influenced mobile application adoption. These factors are supported by a study showing the challenges for adopting technology for diabetes self-management in low-resource areas in the Western Cape, South Africa (Petersen et al., 2019). Perceived behavioural control was also supported by a study that identified the usability of diabetes applications as an important factor in South Africa (Mainoti, Isabirye, & Cilliers, 2019).

Respondents preferred to follow the traditional way of seeking health care advice, uphold traditions and were sceptical about change. The evidence supports Hofstede' view of White South African culture with a low long-term orientation (Hofstede, 2019). The power distance identified in this research may be beneficial when healthcare practitioners recommend and encourage patients to use mobile health applications (Mainoti et al., 2019).

White South Africans form part of an indulgence culture (Hofstede, 2019). They are free to make their own choices. This is consistent with the research finding where diabetic patients can make their own decision whether to use mobile applications for diabetes self-management. The finding of the link between long-term orientation and perceived behavioural control is supported by a study that indicates long-term orientation is positively related to perceived usefulness and perceived ease of use (Sunny et al., 2019).

## 7. CONCLUSION

This study investigated how culture influenced mobile application adoption amongst diabetic patients. Drawing from the research, Hofstede's cultural dimensions and the TPB were used to design the research model. The research contributes to the body of knowledge where there is limited research in this context. The results did not indicate that Hofstede's cultural dimensions and the TPB can identify how culture influences mobile application adoption.

The research can be used to design more culturally sensitive interventions for the context where the use of mobile applications is low. Future research could ask more specific questions, using a cultural model, as it appears that the definition of culture may not be consistent across different groups.

The findings identified in this study are limited to patients with type 2 diabetes residing in the Cape Flats and thus is not a representation of the entire Western Cape population.